# Beyond Resolution: Multi - Scale Weather and Climate Data for Alpine Renewable Energy in the Digital Twin Era - First Evaluations and Recommendations

*Irene Schicker\*(a), Marianne Bügelmayer - Blaschek(b), Annemarie Lexer (a), Katharina Baier (b), Kristofer Hasel (b), Paolo Gazzaneo (b)*

(a) GeoSphere Austria, Hohe Warte 18, A - 1190 Vienna, Austria
(b) Austrian Institute of Technology, Giefinggasse 6, A - 1210, Austria

*Corresponding author: [irene.schicker@geosphere.at](irene.schicker@geosphere.at)*

## ABSTRACT

When Austrian hydropower production plummeted by 44% in early 2025 due to reduced snowpack, it exposed a critical vulnerability: standard meteorological and climatological datasets systematically fail in mountain regions that hold untapped renewable potential. This *perspectives paper* evaluates emerging solutions to the Alpine energy-climate data gap, analyzing datasets from global reanalyses (ERA5, 31 km) to kilometre-scale Digital Twins (Climate DT, Extremes DT, 4.4 km), regional reanalyses (ARA, 2.5 km), and next-generation AI weather prediction models (AIFS, 31 km). The multi-resolution assessment reveals that no single dataset excels universally: coarse reanalyses provide essential climatologies but miss valley-scale processes, while Digital Twins resolve Alpine dynamics yet remain computationally demanding. Effective energy planning therefore requires strategic dataset combinations validated against energy-relevant indices such as population-weighted extremes, wind-gust return periods, and Alpine-adjusted storm thresholds. A key frontier is sub-hourly (10–15 min) temporal resolution to match grid-operation needs. Six evidence-based recommendations outline pathways for bridging spatial and temporal scales. As renewable deployment expands globally into complex terrain, the Alpine region offers transferable perspectives for tackling identical forecasting and climate analysis challenges in mountainous regions worldwide.

# 1. Introduction: The Alpine Energy - Climate Nexus

When Austrian hydropower production plummeted 44% in early 2025 due to reduced snow cover, it wasn't just a regional energy crisis - it was a preview of challenges renewable energy systems face worldwide as deployment expands into complex terrain (Energyquantified, 2025). Similar 31% declines during the 2022 European drought (Copernicus Climate Change Service, 2023) highlight a central paradox of the global energy transition: mountain regions hold abundant renewable and water resources yet pose the greatest challenges for meteorological prediction and climate-risk assessment.

The Alps exemplify this paradox. Extending across eight countries, they offer strong ridge-line winds, high solar irradiation from elevation and snow albedo, and vast hydropower potential from steep gradients and reliable precipitation (Mockert et al., 2023; Dujardin et al., 2021; Schaefli et al., 2019; Gerighausen et al., 2024; Fischer et al., 2025). Yet the very topography that enhances these resources degrades the accuracy of meteorological and climate datasets essential for energy planning - creating an *Alpine energy-climate data gap*.

## 1.1. The Alpine Laboratory for Global Challenges

Recent attribution studies show how climate change amplifies extreme-weather impacts in mountains, with direct consequences for energy systems. The 2022 compound drought and heatwave, which cut Alpine hydropower to 25-year lows, was 3–5 times more likely due to human-induced warming, with >30% of its intensity attributable to anthropogenic factors (Bevacqua et al., 2024; Philip et al., 2023; World Weather Attribution, 2022; Tripathy et al., 2023). These findings turn "exceptional" events into previews of future operating conditions, underscoring the urgency of tighter climate-energy integration in complex terrain.

The stakes extend globally. As renewable deployment accelerates into mountains worldwide, planners find that standard meteorological datasets systematically fail in complex terrain (Lundquist et al., 2019; Poujol et al., 2025; Jimenez et al., 2024). Consequently, methods and data innovations developed for the Alps form critical testbeds for renewable energy expansion in other mountainous regions (Cantao et al., 2023). The relevance is greatest for developing countries, where ≈70% of renewable potential lies in complex topography (RMI, 2024; Singh, 2024; IRENA, 2024). Alpine data solutions could thus inform energy security strategies across the Global South, where renewables are expected to supply 85% of electricity-demand growth through 2035 (World Economic Forum, 2025).

## 1.2. Complex Terrain: Where Climate Models Break Down

Standard global climate datasets (25–150 km resolution) cannot resolve the valley-scale processes dominating Alpine meteorology (Gutmann et al., 2018; Prein et al., 2015). Smoothing of elevation leads to misrepresented wind systems and missing key phenomena such as inversions, föhn events, and orographic precipitation processes that directly influence renewable production (Schicker et al., 2023; Jourdier, 2020; Gruber et al., 2021). Even kilometre-scale regional models show mixed performance depending on parameterization (Poujol et al., 2025).

Across the Alps, ERA5 (Hersbach et al., 2020) - the global "gold standard" - underestimates mountain wind speeds (Jourdier, 2020; Schicker et al., 2023) and omits temperature inversions that can persist for weeks and critically affect both heating demand and solar output. These are not merely technical deficiencies but emerging *energy-security vulnerabilities.*

Wind assessments based on coarse data can overestimate generation by 20–30% in valleys dominated by katabatic flows, while solar estimates neglect fog and low-cloud persistence, overrating winter production (Dujardin et al., 2017). Hydropower projections similarly misrepresent orographic precipitation, underestimating both seasonal variability and extreme-event risks (Wechsler et al., 2023). Glacial retreat, precipitation deficits, and regulatory limits further constrain available water (Wasti et al., 2022).

## 1.3. Toward Energy-Climate Integration in Complex Terrain

Progress requires co-designed climate services that meet energy-sector needs rather than post-hoc adaptation to existing datasets. This is especially vital in the Alps, where infrastructure lifetimes span decades while regional warming proceeds at twice the global rate (Copernicus, 2023). Copernicus now provides derived energy products such as the Pan-European Climate Database (PECD v4.2; Copernicus Climate Change Service, 2024), yet our focus lies on the underlying reanalysis and emerging high-resolution datasets that can capture meteorological extremes and their relevance for renewable systems.

This challenge matters not only for European energy security but for all mountainous regions that hold large untapped renewable potential and host vulnerable populations. The following sections describe current data limitations, examine recent extreme-event case studies, review emerging modelling and Digital Twin solutions, and identify research priorities for enabling reliable renewable-energy deployment in complex terrain worldwide.

# 2. Complex Terrain: Where Standard Datasets Fail

## 2.1 The Orographic Challenges

Mountain meteorology differs fundamentally from flat-terrain dynamics due to three-dimensional atmospheric flow over steep topography (Whiteman, 2000; Serafin et al., 2018; Schicker et al., 2016). The Alps' deep valleys and sharp gradients generate microclimates that strongly affect renewable energy resources but remain unresolved by global climate models.

*Valley winds,* among the most predictable Alpine phenomena, are still underrepresented in standard datasets. These thermally driven circulations - upslope and up-valley by day, downslope and down-valley by night - typically reach 2–5 m s$^{-1}$ (Wagner, 1938; Defant, 1949; Wagner et al., 2015; Serafin et al., 2023). Yet simulations coarser than ~1 km miss these patterns entirely, with biases persisting even at 550 m resolution (Schmidli and Quimbayo-Duarte, 2023).

*Föhn* winds illustrate the dual nature of Alpine flow: warm downslope gusts exceeding 40 m s$^{-1}$ can both supply energy and damage infrastructure (Gohm et al., 2004; Mayr et al., 2007). Their frequency and intensity remain underestimated because orographic triggers occur below ~25 km scales (Jenkner et al., 2010; Drechsel and Mayr, 2008).

*Temperature inversions* create persistent winter microclimates absent from coarse datasets. These stable layers - occurring up to 35% of the time between November and February - reduce valley solar input by 30–50% while enhancing resources above the inversion (Rotach et al., 2004; Largeron and Staquet, 2016; Lehner and Rotach, 2018; Dujardin et al., 2017).

*Terrain shading* further complicates solar assessment: Alpine valleys often receive only 2–4 hours of winter sunlight. Satellite-based irradiance products overestimate valley insolation by >100 W m$^{-2}$, especially under fog and cloud (Hoch and Whiteman, 2010; Solargis, 2024; Li et al., 2024).

## 2.2 Quantifying the Data-Reality Gap

Systematic evaluation of meteorological dataset performance across the Alpine region reveals quantifiable biases that highlight the fundamental challenges of applying global climate data to renewable energy applications in complex terrain. The magnitude of these biases often exceeds acceptable signal-to-noise ratios for energy sector applications.

ERA5 (Hersbach et al., 2020) performs well in flat terrain but loses skill in mountains (Jourdier, 2020; Schicker et al., 2023), underestimating wind speeds and missing inversion dynamics, with temperature RMSE > 3°C in valleys. Regional reanalyses mitigate some errors. CERRA achieves correlations of 0.75–0.82 for wind and temperature (Schimanke et al., 2021; Patra et al., 2025), and COSMO-REA6 up to 0.84 (Bollmeyer et al., 2015; Jourdier, 2020). Yet accuracy declines with elevation (Yang et al., 2025). Typical winter warm biases increase 0.5–1°C per 1.000 m, reflecting unresolved snow-surface feedback, while wind biases shift from -1.2 ms$^{-1}$ in valleys to +0.8 ms$^{-1}$ on ridges.

*Extreme events* reveal the steepest shortcomings. ERA5's 99th-percentile winds correspond only to the observed 90th percentile, implying a 1.5–2 times underestimation of gusts, and the 2003 heatwave's peak temperatures are 2–4°C too low in Alpine valleys. Solar-resource estimates show ±4–8% uncertainty in high-relief regions, mainly due to coarse satellite resolution and unresolved shading, which inflate winter production estimates (Solargis, 2024).

Overall, biases scale with terrain complexity, underscoring that kilometre-scale resolution alone is insufficient without improved parameterizations of orographic processes and boundary-layer physics.

## 2.3 Infrastructure Vulnerability Multipliers

The systematic limitations of meteorological datasets in complex terrain create cascade effects that amplify infrastructure vulnerabilities beyond the direct impacts of individual weather events. Alpine energy systems operate in isolated networks with limited redundancy, where meteorological prediction errors compound through interconnected infrastructure systems. Access to Alpine infrastructure is often limited during adverse weather, amplifying vulnerabilities when forecasts are inaccurate. Mountain roads, cable routes, and helicopter access can be blocked for days, delaying repairs when föhn events damage transmission lines in remote areas.

Forecasting deficiencies also hinder grid integration. Poorly resolved valley-wind and inversion patterns cause multiple wind farms to under- or over-perform simultaneously, creating regional balancing challenges (Serafin et al., 2018). Because standard datasets miss the coherent meteorological structures linking sites, grid operators must compensate with larger reserves and conservative operating strategies - reducing the economic competitiveness of renewables in complex terrain.

These weaknesses create cascade effects across interconnected systems: a single föhn event may damage lines while accelerating snowmelt, compounding risks that exceed component design limits. Hydropower across mountain regions will face greater sediment fluxes in some seasons and reduced flows in others as competing demands intensify (European Environment Agency, 2025).

Climate change further magnifies these stresses. Extreme-weather damages already exceed €790 billion and 246,000 fatalities in Europe since 1980 (European Environment Agency, 2023), and without adaptation, losses are expected to rise sharply (IFAB, 2023). These challenges extend beyond infrastructure to markets and policy frameworks that still rely on stationary-climate assumptions.

The systematic limitations of meteorological datasets amplify infrastructure vulnerabilities through cascade effects that propagate across interconnected Alpine energy systems. When föhn wind events damage transmission infrastructure while simultaneously accelerating snowmelt, the compound risks create system-wide vulnerabilities that exceed individual component design criteria. Hydropower in all mountain regions will experience higher water and sediment fluxes in some seasons but lower water flow with competing demands in others (European Environment Agency, 2025). Climate change and extreme weather events increasingly affect all parts of the European energy system, with direct damages potentially amounting to significant economic losses without appropriate adaptation measures (IFAB, 2023). The integration challenges extend beyond physical infrastructure to market and policy domains where energy markets operate on probabilistic forecasts assuming stationary climate conditions, yet extreme weather has been responsible for around €790 billion in economic losses and approximately 246.000 human fatalities across Europe from 1980 to 2023 (European Environment Agency, 2023).

## 2.4 Compound Extremes Amplify Energy System Vulnerabilities

Compound extreme events - where multiple hazards occur simultaneously - pose an escalating challenge for Alpine energy systems that current meteorological datasets cannot adequately capture. Such events create cascading impacts exceeding the sum of individual drivers and expose core weaknesses in existing energy-planning frameworks (Hao et al., 2022; van der Wiel et al., 2024).

The 2022 European drought-heatwave illustrates this vulnerability: Europe's warmest summer on record (+2 °C above average) coincided with drought across 63% of rivers (Copernicus, 2023; Tripathy et al., 2023). Hydropower capacity factors dropped by 6.5%, cooling demand surged, the Po River fell to 70-year lows, and Alpine glaciers lost over 5 km³ of ice - reducing future seasonal storage (Fernández - González et al., 2023).

Attribution studies show these extremes were 3–5 times more likely due to anthropogenic warming, with >30% of their intensity directly linked to human influence (Bevacqua et al., 2024; Philip et al., 2023). By the late 21st century, about 20% of global land areas may face two such drought-heatwave events annually, each lasting around 25 days (Manning et al., 2023).

These compounding risks highlight the urgent need for high-resolution, physically consistent datasets that can represent multi-hazard interactions - precisely the gap addressed by the emerging solutions discussed in Section 3.

# 3. Emerging Solutions: Bridging the Alpine data gap and the Digital Twins Revolution

## 3.1 Legacy and Current Operational Reanalysis Datasets

Meteorological reanalyses have steadily advanced data quality for complex terrain, yet key limitations persist. ERA5, ECMWF's fifth-generation global reanalysis, remains the operational benchmark, offering hourly data from 1940 onward at 31 km resolution (Hersbach et al., 2020; Soci et al., 2024). It introduced major improvements over ERA-Interim, including higher temporal resolution and a 10-member ensemble (ERA5-EDA) that provides uncertainty estimates. Nevertheless, ERA5's coarse resolution cannot resolve valley-scale Alpine processes. The companion ERA5-Land (surface) product increases spatial detail to ~9 km through offline land-surface downscaling, enhancing near-surface fields and orographic representation (Muñoz - Sabater et al., 2021).

Regional reanalyses bridge the gap between global products and local requirements. The Copernicus European Regional ReAnalysis (CERRA) operates at 5.5 km resolution since 1984, using the HARMONIE-ALADIN model with 3D-Var data assimilation and additional national observations (Ridal et al., 2024). CERRA provides improved surface-variable realism in complex terrain and includes an 11 km ensemble (CERRA-EDA) offering flow-dependent background error estimates. Validation confirms superior skill for wind, temperature, and biometeorological variables in mountainous regions (Galanaki et al., 2023).

The COSMO-REA family, developed by Germany's DWD, extends this progress with COSMO-REA6 (6 km, Europe-wide, 1995–2019) and COSMO-REA2 (2 km, convection-permitting, Central Europe) (Bollmeyer et al., 2015; Kaspar et al., 2020). Both use nudging-based assimilation with dedicated snow and soil-moisture analyses. COSMO-REA6 achieves mean absolute wind-speed errors of 0.69 ms$^{-1}$ - clearly outperforming global reanalyses - although production ended in 2019 following ERA-Interim's discontinuation.

The Austrian Reanalysis (ARA) ensemble (Awan et al., 2024) represents the first reanalysis tailored explicitly to Alpine terrain. Using the non-hydrostatic AROME model at 2.5 km resolution with convection-permitting physics, ARA applies 3D-Var assimilation and C-LAEF ensemble techniques to produce a 10-member plus control run system spanning 2012–2021). By assimilating satellite, radiosonde, aircraft, and wind-profiler data, ARA markedly improves representation of valley-wind systems, temperature inversions, and precipitation structures compared with coarser products.

## 3.2 Convection-Permitting Climate Modelling and CORDEX FPS-Convection

Convection-permitting climate modelling is revolutionizing our ability to represent mountainous region meteorology. The CORDEX Flagship Pilot Study on Convective Phenomena (CORDEX FPS-Convection) represents the first multi-model ensemble of kilometre-scale simulations over the extended Alpine domain, building a benchmark dataset for assessing added value at convection-permitting scales (Coppola et al., 2020). The initiative encompasses hindcast simulations forced by ERA-Interim reanalysis and historical and future climate scenario simulations, with multiple regional climate models operating at 2–3 km resolution over the Alpine region.

Recent studies document substantial added value from CORDEX FPS-Convection simulations. For temperature over complex orography, kilometre-scale simulations demonstrate systematic improvements in representing maximum and minimum temperatures compared to coarser-resolution counterparts (Soares et al., 2023). For near-surface wind speed, the multi-model ensemble shows significant added value over the Alps, with convection-permitting models better capturing the spatial variability and extreme values critical for wind energy applications (Molina et al., 2024). The ensemble approach allows quantification of both model uncertainty and the magnitude of improvements achievable through increased resolution.

In recent years, researchers have pushed computational limits further, advancing into the hectometric scale (100m–1km resolution) entering the grey zone between convection-resolving and large eddy simulations, as in the Destination on-demand digital twin (weather forecasts). These high-resolution simulations can partly resolve or parameterize large eddies associated with up- and downdrafts near clouds, improving understanding of convection and cloud formation processes (Nuissier et al., 2020; Stevens et al., 2020; Schulz and Stevens, 2023). While hectometric scale climate simulations offer advances in representing complex terrain processes, they pose validation challenges due to sparse observations at comparable resolutions, making high-density observation networks and high-resolution reanalysis datasets increasingly critical.

## 3.3 Operational Analysis and Nowcasting Systems

The Integrated Nowcasting through Comprehensive Analysis (INCA) system, developed by GeoSphere Austria (formerly ZAMG), provides high-resolution analyses and short-term forecasts tailored for mountainous terrain (Haiden et al., 2011). INCA merges surface observations with radar and satellite data at 1 km spatial and hourly temporal resolution, reproducing local station values while adding spatial structure from remote sensing. It delivers temperature, humidity, wind, precipitation, cloudiness, and radiation - core variables for renewable-energy applications.

INCA shows particularly strong skill for temperature, with clear added value during the first few forecast hours and continued improvement when dynamically downscaling NWP guidance at longer lead times (Haiden et al., 2011; Ghaemi et al., 2021). For Alpine energy use, it acts as a vital interface between coarse global models and the detailed meteorological information required for operational decisions. The system underpins flood forecasting, winter maintenance, and public weather services across Austria.

Looking ahead, hybrid AI-physics approaches such as the Anemoi framework (Prieto Nemesio et al., 2025) - while currently being used for short to medium range predictions with first experiments for nowcasting - offer a promising next step. Anemoi integrates machine-learning inference with conventional modelling, enabling kilometre-scale regional forecasts updated in near-real time. Its architecture could eventually extend INCA - like nowcasting toward rapid, AI-driven ensemble generation for localized renewable-energy forecasting and extreme-event monitoring.

## 3.4 Next-Generation Forecasting: Physics-Based and AI-Driven Approaches

The ECMWF Integrated Forecasting System (IFS) remains the operational backbone of European medium-range prediction, advancing through regular cycle upgrades. The high-resolution configuration (HRES) runs at 9 km with 137 levels, while the 51-member ensemble (ENS) reached the same 9 km grid with Cycle 48r1 (June 2023) (Haiden and Chevallier, 2025). This represents the highest-resolution global forecasting system currently in operation and provides key boundary conditions for regional models across ECMWF Member States. IFS employs advanced 4D-Var assimilation, ingesting ~60 million satellite, aircraft, radiosonde, and surface observations every six hours. For Alpine applications, it delivers the large-scale forcing and validation reference for emerging AI-based systems.

The Artificial Intelligence Forecasting System (AIFS), operational since February 2025, marks a milestone as the first fully operational AI-based global model (Lang et al., 2024). Built on a graph-neural-network encoder-processor-decoder architecture trained on ERA5 and IFS analyses, AIFS achieves comparable or superior skill to physics-based models for many variables. It generates global 10-day forecasts in minutes on GPUs - about 1.000 times more energy-efficient than conventional systems - and its 51-member ensemble (AIFS-ENS) became operational in July 2025 (ECMWF, 2025). For Alpine energy forecasting, AIFS offers the speed and scalability needed for rapid ensemble generation during extreme events.

Beyond global models, regional AI forecasting is rapidly evolving. Met Norway's "Bris" system applies a stretched-grid architecture with 2.5 km resolution over the Nordic region while retaining coarser (~100 km) global coverage (Nipen et al., 2024). Trained on ERA5 and MetCoOp ensemble data, Bris delivers 10-day forecasts in 2–3 minutes on a single GPU and already outperforms operational physics-based systems for temperature (EuroHPC, 2025).

This stretched-grid concept underpins the Anemoi framework, a collaborative open-source initiative led by ECMWF and European partners to develop regionally focused AI forecasting systems (Prieto Nemesio et al., 2025). By combining AI inference with traditional modelling, Anemoi provides a blueprint for future kilometre-scale, AI-driven nowcasting and regional prediction, directly supporting renewable-energy and extreme-event applications in complex Alpine terrain.

## 3.5 Digital Twins and Kilometre-Scale Earth System Modelling

The European Union's Destination Earth (DestinE) initiative represents a fundamental shift toward operationally relevant, high-resolution Earth system modelling (ECMWF, 2024. Hoffmann et al., 2023). DestinE comprises two complementary digital twins addressing critical Alpine energy-climate challenges:

The Weather-Induced Extremes Digital Twin (Extremes DT) combines a global continuous component at 4.4 km resolution with an on-demand regional component achieving 500–750 m resolution over Europe. The on-demand system, developed by Météo-France with 22 European partners, provides configurable simulations activated when extreme events are predicted, finally achieving spatial scales sufficient to resolve individual Alpine valleys and local orographic processes (Hoffmann et al., 2023). This capability enables unprecedented impact assessment for renewable energy infrastructure during extreme weather events.

The Climate Change Adaptation Digital Twin (Climate DT), implemented by Finland's CSC-IT Centre for Science with partners across six countries, provides multi-decadal global climate simulations at approximately 4–9 km resolution. The Climate DT employs innovative data streaming concepts, allowing impact-sector applications to access information during simulation production rather than post-processing stored data (ECMWF 2024b). For renewable energy applications, the Climate DT enables assessment of future extreme event frequency and provides boundary conditions for hectometric-scale simulations of Alpine energy systems under climate change.

*Table 1: A summary of reanalysis datasets, forecasting systems, and Digital Twin data for Alpine Energy Applications. It is, however, out of the scope of this perspective paper to discuss all datasets. Datasets denoted with a * are discussed in more detail.*

| Dataset/System | Horizontal Resolution | Temporal Coverage | Update Frequency | Key Technology | Key Features for Complex Terrain |
|---|---|---|---|---|---|
| ERA5* | ~31 km | 1940–present | Daily updates | 4D-Var | Global coverage; 10-member ensemble; operational standard |
| ERA5-Land* | ~9 km | 1950–present | Daily updates | Offline land model | Enhanced land-atmosphere; improved topography |
| CERRA | ~5.5 km | 1984–present | Updates planned | 3D-Var (HARMONIE) | Regional Europe; local observations; Alpine validation |
| COSMO - REA6 | ~6 km | 1995–2019 | Discontinued | Nudging | Energy sector validation; Alpine focus |
| ARA* | ~2.5 km | 2012–2021 | Research | 3D-Var (AROME) | Convection-permitting; Austria-specific; 10-member ensemble, one control run |
| INCA | ~1 km | 2007–present | Hourly | Analysis/nowcast | Real-time; mountainous terrain; blended observations |
| CORDEX FPS - Conv | ~2–3 km | Hindcasts/scenarios | Research | Multi-model RCMs | Convection-permitting; multi-model; Alpine domain |
| ECMWF IFS HRES | ~9 km | Operational | 6-hourly (10-day) | 4D-Var physics | Highest-resolution global operational model |
| ECMWF IFS ENS | ~9 km | Operational | 6/12-hourly (15-day) | 4D-Var physics | 51-member ensemble; probabilistic |
| AIFS Single* | ~31 km | Operational | 6-hourly (10-day) | GNN machine learning | 1000 times faster; energy efficient; operational AI |
| AIFS ENS | ~31 km | Operational | 6/12-hourly (15-day) | GNN + diffusion | 51-member AI ensemble; probabilistic |
| Bris (Met Norway) | ~2.5 km | Semi-operational | 6-hourly (10-day) | Stretched-grid GNN | Regional high-resolution; seamless nesting; rapid inference |
| Extremes DT* (Continuous) | ~4.4 km | Operational (forecast) | 6-hourly | Coupled physics | Near-real-time extremes; Europe-wide |
| Extremes DT (On-demand) | ~500–750 m | Operational (on-demand) | Event-triggered | Multi-model ensemble | Sub-km simulations; valley-resolving; configurable |
| Climate DT* | ~4.4–9 km | Multi-decadal projections | Scenario-based | Coupled ESM (ICON/IFS) | Climate scenarios; data streaming; sectoral apps |
| NextGEMS | ~5–10 km | 30-year projections | Research | ICON/IFS-FESOM ESM | Storm-resolving; foundation for Climate DT |

These digital twins build upon nextGEMS (Next Generation Earth Modeling Systems), a Horizon 2020 project that achieved the first multi-decadal (30-year) climate simulations at kilometre scales using the ICON and IFS-FESOM Earth system models. Operating at approximately 5 km ocean resolution and 10 km atmosphere resolution, nextGEMS simulations achieved a throughput of about 500 simulated days per day on the Levante supercomputer, providing unprecedented realism in representing convective processes, orographic effects, and ocean eddies (Segura et al., 2025). NextGEMS models form the scientific foundation of the DestinE Climate DT and pave the way for future European research on climate change impacts at scales directly relevant to energy infrastructure planning.

The convergence of convection-permitting regional models, operational digital twins, AI-driven forecasting, and energy sector co-design creates unprecedented opportunities to bridge the Alpine data-reality gap. For the first time, energy planners can access meteorological datasets and forecasts at scales matching infrastructure deployment decisions, with explicit representation of the valley-scale processes that dominate Alpine renewable resource variability. The integration of physics-based and AI-driven approaches - with IFS providing physical consistency and AIFS offering computational efficiency - enables both rapid ensemble generation for operational decision-making and long-term climate projections for strategic planning.

# 4. Bridging Weather and Climate: From Historical Assessment to Future Projections - Multi-Resolution Dataset Evaluation

Renewable-energy planning requires seamless integration across temporal scales: reanalysis data establish historical resource climatologies and extreme event frequencies, forecasting systems enable operational risk management and infrastructure stress testing, while climate projections inform long-term investment and adaptation strategies. However, these three domains - historical assessment, weather prediction, and climate projection - have traditionally operated independently, using different models, resolutions, and evaluation frameworks. Recent advances in Digital Twin technology and high-performance computing now enable unprecedented continuity: the same model physics applied at weather timescales (hourly, days) and climate timescales (monthly, decades), at resolutions (≤5 km) that (hopefully) match Alpine/complex terrain renewable energy and infrastructure deployment scales. This section evaluates some of the datasets mentioned in Section 3 spanning this weather-to-climate continuum, demonstrating how spatial and temporal resolution fundamentally determine their utility for Alpine renewable energy applications. We progress from coarse global reanalysis establishing multi-decadal baselines, though do not look deeper into hectometric forecasts resolving individual valleys and extreme events, to preliminary climate projections at unprecedented resolution. This multi-scale approach reveals that meaningful Alpine energy assessment requires not just high resolution OR long temporal coverage, but both simultaneously- a requirement only now becoming technically feasible. While, thus, a complete evaluation of all datasets in Table 1 exceeds the scope of this perspectives paper, we present representative results that demonstrate the usability, skills, and results of the selected datasets in terms of renewable energy applications decisions in semi-complex and complex Alpine terrain. In the following, the datasets used are briefly described, the overall implemented type of events, our "evaluation framework" and energy-alpine related phenomena discussed. A subset of the results is presented but the data will be made available on Zenodo and the code is available on the GitHub page (https://github.com/ischicker/perspectives_iScience_ClimateEnergyNexus_2025).

## 4.1 Datasets used - Overview

This study evaluates five state-of-the-art reanalysis and high-resolution modelling datasets for their suitability in Alpine renewable energy applications: ERA5 (31 km, hourly), ERA5-Land (9 km, hourly), ARA (2.5 km, hourly), and the Climate DT (4.4 km, hourly). These datasets represent a spectrum of spatial resolutions and modelling approaches specifically relevant to complex terrain applications. More details on the datasets can be found in Table 1.

## 4.2 Evaluation Framework and Analysed Phenomena

A set of energy, alpine, and energy-alpine specific related phenomena including adaptable thresholds were implemented to both highlight the alpine data and resolution gap and the skills of the novel and upcoming datasets. An overview of the most important phenomena investigted is given in Table 2. This includes, besides thresholds, a description of the phenomena and their relevance in terms of energy and alpine specific needs/necessities.

As it is out of the scope of this perspectives paper to discuss all results and to evaluate a plethora of data, results of selected phenomena and data sets are briefly discussed.

*Table 2: A brief overview of the implemented evaluation methods, thresholds, description, and key references.*

| Evaluation Method | Thresholds and Parameters | Description and relevance (energy, alpine specific) | Key References |
|---|---|---|---|
| **Wind Gust Analysis** | Strong: 20 m/s<br>Severe: 25 m/s<br>Extreme: 32.7 m/s (ESSL) | Structural loading exceeds mean wind effects; gust factors amplified in complex terrain. Turbine cut-out events, structural design, downtime estimation, insurance assessment. Topographic channeling and flow separation create localized extreme gusts; valley-ridge gust factor variation up to 50%. | Wieringa (1973); Schicker and Meirold-Mautner (2022); IEC 61400-1 |
| **Population-Weighted Temperature** | Heat: >30°C<br>Cold: <−10°C<br>Population density overlay | Meteorological accuracy matters most where people live; energy demand concentrated in populated areas.<br>Heating/cooling degree days, demand forecasting, grid capacity planning. | van der Wiel et al. (2020); Schiavina et al. (2023) |

| | | Extreme population-climate decoupling in the Alps; valley cities vs. unpopulated peaks; concentrated heating demand. | |
|---|---|---|---|
| **Dunkelflaute Events** | Wind: <3 m/s  Solar: <50 W/m²  Duration: ≥24 h | Low renewable supply periods drive backup capacity needs and energy storage requirements.  Backup generation dispatch, battery sizing, interconnection planning, and reliability analysis.  Winter Alpine inversions create persistent low-wind, low-solar conditions; multi-day events common | Ohlendorf and Schill (2020); Bloomfield et al. (2021) |
| **Storm Classification (ESSL-based)** | Standard ESSL thresholds:  + Severe gusts: ≥25 m/s  + Extreme gusts: ≥32.7 m/s  + Alpine-adjusted thresholds:  + Severe: ≥20 m/s  + Extreme: ≥28 m/s - (15–20% terrain reduction) | European Severe Storms Laboratory severity standards with Alpine-specific adjustments necessitated by complex terrain interactions. Severe gusts trigger turbine cut-out mechanisms and emergency shutdowns halting energy production. Extreme gusts pose structural hazards requiring extended maintenance inspections. Alpine threshold reductions reflect empirical findings that complex topography intensifies local flow acceleration and turbulence through valley channeling, gap flows, and lee-side effects - reaching equivalent damage potential at lower baseline wind speeds. Multi-tier classification quantifies both productivity losses (hours offline) and infrastructure risk (structural loading frequency). | Wieringa (1973); Schicker and Meirold-Mautner (2022); IEC 61400-1; ESSL criteria |
| **Heating/Cooling Degree Days** | HDD base: 15°C/18°C  CDD base: 18°C/22°C  Daily aggregation | Direct proxy for temperature-driven energy demand; widely used in the energy sector.  Demand forecasting, seasonal energy planning, and building efficiency assessment.  Extreme altitude-dependent HDD variation; valley inversions affect CDD; Föhn events | Matzarakis and Thomsen (2009); Spinoni et al. (2015) |
| **Compound Extremes** | Heat (>30°C) + Drought (<20% precip Norm)  Wind drought + Solar drought  Duration: ≥72 h | Coincident stressors exceed impacts of individual events; energy systems vulnerable to compound stress.  Grid stress events, renewable complementarity assessment, resilience planning.  Orographic effects create spatial compound event patterns; blocking situations affect the entire Alpine arc. | Zscheischler et al. (2020); van der Wiel et al. (2020) |
| **Hydropower Resource Assessment** | Heavy precipitation: >50 mm/day  Alpine runoff coefficient: ~40%  Resource potential:  + High: >3 mm/day runoff  + Good: 2–3 mm/day  + Moderate: 1–2 mm/day | Precipitation/runoff climatology for hydropower potential assessment critical to Alpine energy systems. Seasonal patterns driven by snowmelt plus precipitation create distinct resource availability. Orographic precipitation enhancement on windward slopes (2–5 times amplification) creates sharp spatial gradients requiring high-resolution datasets - coarse products typically underestimate terrain-induced modification. | Frei and Schär (1998); Dubus et al. (2023); Marty et al. (2002) |

Essential for dam operations, flood risk assessment, reservoir management, and annual energy planning. Complex terrain effects poorly captured by standard reanalysis products.

*Historical Foundation: Reanalysis Datasets - New horizons with ensemble reanalysis*

In the past years, several initiatives on generating high resolved reanalysis data sets started (Awan et al.; 2024, Yang et al., 2025; Bazile et al., 2025) including the formation a new WEMC reanalysis working group (see https://www.wemcouncil.org/wp/reanalysis-working-group/ for more details). Beyond data availability considerations, we focus our analysis on 2020, a year that exemplified the energy-climate challenges ahead: it ranked among the three warmest on record globally (WMO, 2021), experienced US$210 billion in natural disaster losses (Munich Re, 2021), and featured Storm Alex which broke precipitation records across the Alps (Copernicus, 2021) while demonstrating extreme hydropower variability - precisely the compound energy-climate extremes that high-resolution reanalyses must capture.

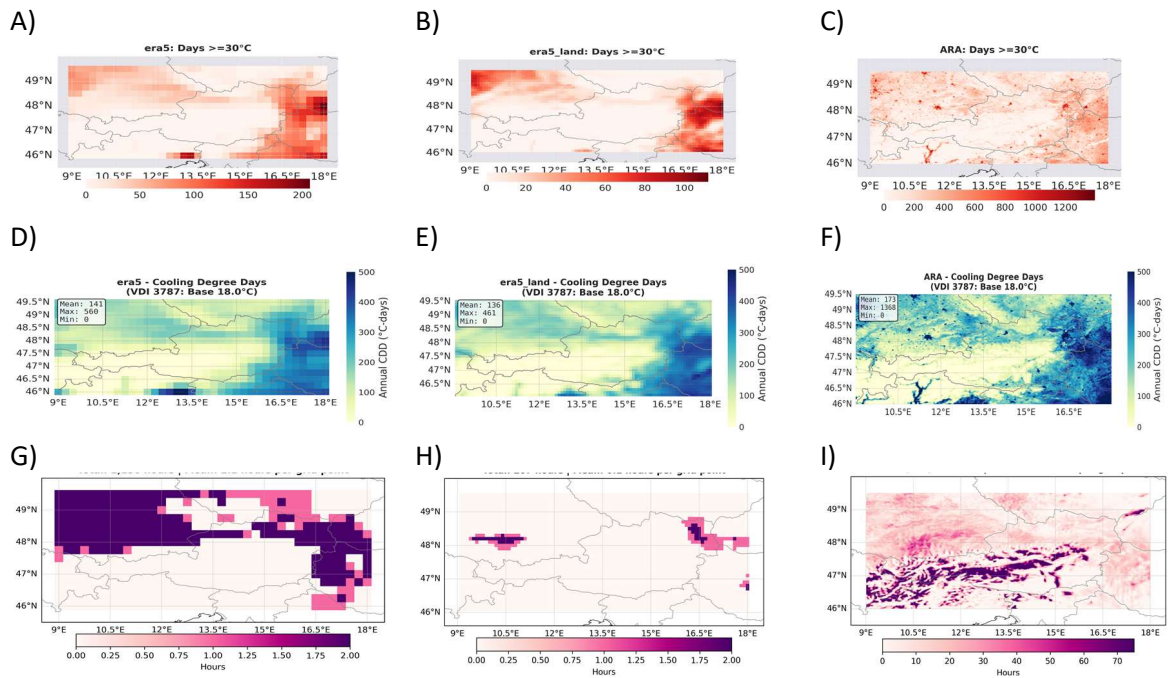

*Figure 1: Comparison of ERA5, ERA5-Land and ARA (control member) for the year 2020. The resolution impact on Alpine and energy related indices is clearly distinguishable. Top row: number of population-weighted temperature days with temperatures >30°C reveal increasingly fine-scale thermal structure as resolution improves from ERA5's 31 km (A) through ERA5-Land's 9 km (B) to ARA's 2.5 km convection-permitting resolution (C), with valley-bottom cooling demand hotspots invisible in coarser products. Middle row: Cooling Degree Days (base 18°C) for ERA5 (D), ERA5-Land (E), and AFA (F). Bottom row: storm days with mean wind speed >17.5 m/s frequency maps show ARA's (I) ability to resolve both the topographical related regions and the detailed structures needed for energy planning, systematically underrepresented in globally gridded reanalyses.*

Results for 2020 (Figure 1) demonstrate the impact of spatial resolution on Alpine, or in general more complex topographical regions, energy assessments for ERA5 (31 km), ERA5-Land (9 km), and ARA (2.5 km) data. Examining three energy-critical indicators, population-weighted heat extremes (>30°C) that drive peak electricity demand and cooling loads during concurrent heat stress across populated regions; cooling degree days (CDD) essential for sizing air conditioning systems and district cooling networks; and storm days (mean wind ≥17.5 m/s) that determine wind turbine survival requirements, transmission line failure risk, and emergency shutdown protocols, ARA's convection-permitting resolution exposes localized patterns in all parameters investigated (see Table 2). Revealing micro-regional hotspots where population exposure, cooling energy demand, and infrastructure vulnerability intersect- that are systematically averaged away in coarser reanalyses, demonstrating that sub-5 km resolution is non-negotiable for identifying where climate extremes and energy system stress co-occur. Besides using reanalysis like ARA as methodological template for future operational Alpine reanalyses and climate services, such data sets build the base for stretched-grid AIWP models such as Bris (Nipen et al., 2024). They also prove that convection-permitting systems can deliver the valley-scale precision essential for municipal energy planning, grid reinforcement decisions, and targeted climate adaptation investments. These capabilities will become increasingly critical as the (Alpine) energy transition accelerates and extreme event frequencies intensify under climate change.

*Operational Bridge: High-Resolution Forecasting (hours to days) - From Climatology to Specific Events*

Storm Benjamin hit western and Central Europe on October 23rd, 2025, with widespread gusts exceeding 120 kmh⁻¹, heavy rainfall, a high coastal surge risk, and localized tornadic activity. Its rapid evolution and small-scale wind maxima, including a tornado north of Paris (The Times, 2025), exposed the persistent resolution gap between global and regional forecasting systems. It is used here as an example for bridging the gap between weather and climate and impacts on the energy sector. With the resolution of the Extremes DT of 4.4 km, which is also shared by the Climate DT, it enables a more detailed view on the impact of such systems. Especially, with a projected intensification under a warming climate (Harvey et al., 2020). For the energy sector, Benjamin highlighted the vulnerability of wind and grid infrastructure to short-lived but intense gust events and the need for high-resolution, ensemble-based prediction and impact modelling.

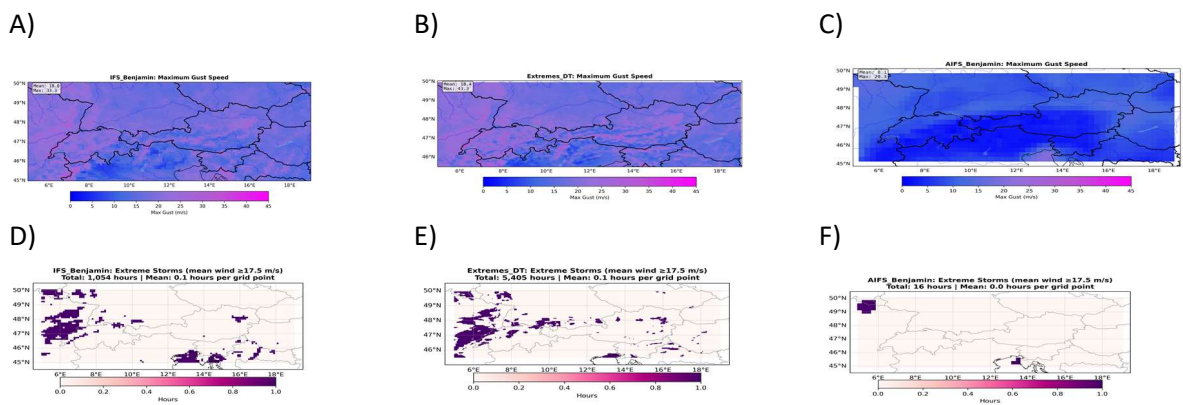

*Figure 2: Comparison of top row: maximum 10-m wind gusts during Storm Benjamin (October 23, 2025, 12:00 UTC) from (A) ECMWF IFS operational forecast (9 km resolution), (B) Extremes Digital Twin high-resolution simulation (4.4 km), and (C) AIFS AI-based forecast (31 km). Bottom row: severe storm wind speed values for (D) IFS, (E) Extremes DT, and (F) AIFS with red contours indicating wind thresholds (15 m/s).*

A brief evaluation of the different skills of the models IFS, Extremes DT, and AIFS during the storm event (Figure 2) shows that the Extremes DT's enhanced resolution captures fine-scale wind gust maxima exceeding 15 m/s and mesoscale features as well as partially higher intensities in max. gust speeds and duration. AIFS, currently operating at coarser resolution limiting its ability to resolve valley-scale extremes, while underestimating wind speed and intensities but capturing the tendencies, still offers a promising potential for the future operational decision support particularly with a migration to higher spatial and temporal resolutions soon.

*Future Projections: Climate at Weather-Relevant Scales - Climate Digital Twins - Unprecedented Resolution for Adaptation Planning*

With the Destination Earth Climate DT simulations at 4.4 km resolution demonstrating, like the Extremes DT, unprecedented capability to resolve valley-scale energy phenomena, critical for adaptation planning, future renewable energy related planning and/or mitigation and adaptation planning can be made more accurately and precise. These capabilities build upon the nextGEMS project, which demonstrated the first multi-decadal kilometre-scale climate simulations and proved that sub-10 km Earth system modelling is computationally viable for long-term climate projections. For this perspectives paper, though our analysis covers January-March 2020 due to computational constraints and data availability limitations and the full year 2022 (Figure 3). Even within this limited 3-month winter window, the Climate DT IFS-based scenario reveals spatially heterogeneous patterns in heating degree days and severe wind frequency (mean wind ≥15 m/s) that expose localized infrastructure vulnerabilities invisible in traditional 30 km climate projections. Similar findings can be drawn for the 2022 data. Particularly, the 2–3 times HDD variation between adjacent valleys that fundamentally alters heating system design requirements and the wind corridor identification essential for both turbine siting and transmission line risk assessment. This resolution breakthrough enables the first site-specific, municipal-level adaptation planning where energy infrastructure deployment occurs, with full multi-decadal Climate DT datasets expected to transform long-term energy system resilience planning as computational resources and data archives expand.

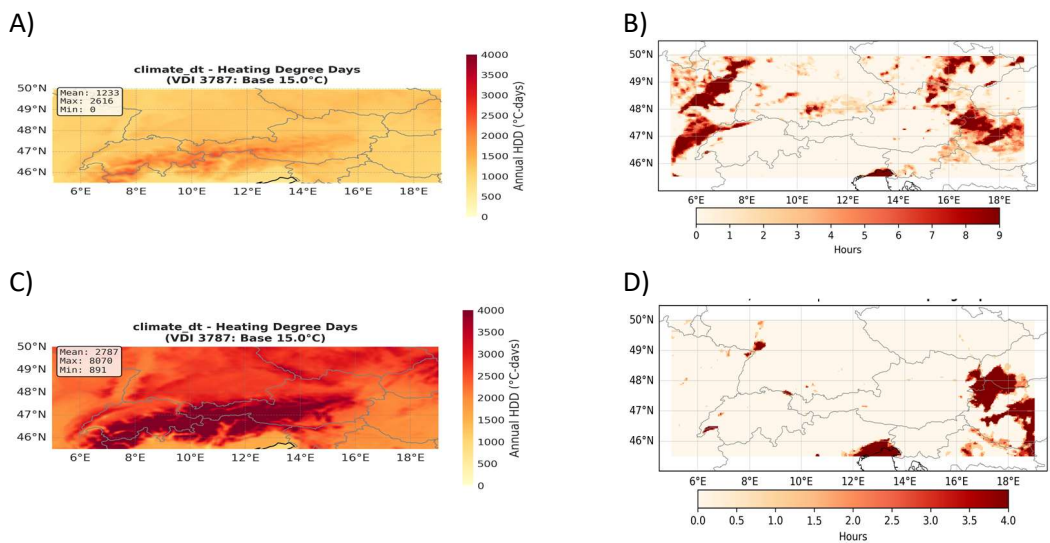

*Figure 3: Top row: results for the months January to March 2020 for A) Heating degree days (HDD, VDI 3787 base 15.0°C) reveal strong elevation-dependent gradients (mean: 1,233°C-days, max: 2,616) with valley-scale heterogeneity critical for heating infrastructure planning. B) Storm frequency (mean wind ≥15 m/s) enables*

*the localized high-wind corridors identification. Bottom row: evaluation for the Climate DT IFS-based scenario 2022 for C) HDD for the full year and D) storm frequency for the full year.*

# 5. Recommendations and Future Directions

Bridging weather and climate data for (Alpine) renewable energy requires integration across spatial, temporal, and process scales. No single dataset performs optimally everywhere: coarse global reanalyses offer long climatological baselines but miss valley-scale dynamics, while kilometre-scale Digital Twins such as the Climate DT 4.4 km or the Extremes DT capture key (Alpine) processes but remain computationally demanding. Effective planning ideally combines datasets - ERA5 for long-term characterization, regional reanalyses (CERRA, ARA) for operational use, and emerging high-resolution projections for infrastructure adaptation. However, a major remaining gap concerns temporal resolution. Hourly data suffice for resource assessments but not for grid operations that depend on minute-scale variations in wind and solar output. Even 3-hour changes mask finer fluctuations crucial for real-time balancing. Future operational Digital Twins ideally operate with 10–15 min output frequency to align with grid decision cycles, especially in complex Alpine terrain where weather changes rapidly. Closing this temporal-resolution gap will require co-designed data systems, validation frameworks, and joint investment to ensure renewable-driven grids to remain stable and resilient.

To comprehensively and successfully tackle the challenges described above, six key recommendations are formulated:

1. Match Data with Application: Select meteorological datasets based on specific application needs, considering spatial and temporal resolution requirements, parameter availability, and quality characteristics. There is a steep increase in available data related to research focusing on mountainous areas and intensified market needs. Therefore, regular assessment of available data sets and parameters is recommended.
2. Combine Multiple Sources: Integrate information from multiple datasets to overcome individual limitations. For example, combine high-resolution reanalysis with site-specific measurements and specialized downscaling techniques in complex terrain. Further, the institutions within the countries along the Alps are increasingly engaging and aligning the produced data to enhance the representativeness of the Alpine region. An effort that needs further focus and should be deployed globally.
3. Validate Against Independent Observations and Energy-Relevant Indices and Metrics: Rigorously validate datasets against independent measurements, particularly in complex terrain regions where model performance may vary significantly. Benchmark datasets against independent observations using both standard and sector-specific measures: diurnal wind reversals, inversion frequency, orographic precipitation, and föhn detection. However, not only do the model data sets need to be critically examined, but also the observation data requires thorough assessment regarding its quality and its representativeness of the area.
4. Consider Uncertainties: Acknowledge, state, and quantify uncertainties in meteorological data, especially when assessing extreme events or long-term climate impacts. By clearly outlining the limitations and related uncertainties, the applicability of the data set can be ensured, and sound decisions can be taken. Uncertainty quantification is crucial yet still needs further interdisciplinary research when looking at renewable energy planning, as uncertainty stems from climate data, technological, and socio-economic developments.
5. Standardize Validation Methodologies: Adopt, develop, and implement standardized validation protocols with metrics specifically designed for renewable energy applications. This aspect strongly supports the previous point as only standardized approaches can enable a general assessment of uncertainties. International adoption of such standards would ensure consistent evaluation and guide dataset selection for specific renewable-energy applications.
6. Invest in Improved Data Resources: Support the development and maintenance of high-resolution observational networks and reanalysis products, particularly in data-sparse regions and complex terrain.

Emerging operational Digital Twins and AI-driven forecasting systems enable, for the first time, energy-relevant meteorological information at deployment scales. Translating this technical capability into operational benefit requires standardized validation methodologies, systematic uncertainty quantification, and collaborative frameworks bridging meteorological and energy communities. Future work should leverage derived energy products (e.g., PECD v4.2) to translate detected extremes directly into system impacts, while developing sub-hourly forecasting capabilities essential for grid stability in high-penetration renewable systems. The multi-resolution extreme detection and validation framework presented here provides a first meteorological foundation for such energy-specific analyses, with implications extending globally to all mountainous regions undergoing renewable energy transitions.

To conclude, this perspective paper presents a short overview of current challenges and emerging solutions for renewable energy planning within mountainous areas, focusing on the Alps, with respect to meteorological and climatological data. The analyses, spanning coarse reanalysis to emerging kilometer-scale Digital Twins, reveals fundamental trade-offs between temporal coverage and spatial resolution that necessitate application-specific dataset selection and multi-source synthesis approaches.

## AUTHOR CONTRIBUTIONS

<u>Conceptualization</u>: I.S., M.B.-B.; <u>methodology</u>: I.S.; <u>formal analysis</u>: I.S.; <u>data curation</u>: I.S., P.G., K.H.; <u>writing - original draft</u>: I.S., M.B.-B., A.L.; <u>writing - review and editing</u>: all authors; <u>visualization</u>: I.S., K.H.; <u>supervision</u>: I.S.; <u>project administration</u>: M.B.-B., I.S.; <u>funding acquisition</u>: all authors

## DECLARATION OF INTERESTS

The authors declare no competing interests.

## RESOURCE AVAILABILITY

Further information and requests for resources should be directed to and will be fulfilled by the lead contact, Irene Schicker (irene.schicker@geosphere.at).

Materials Availability: This study did not generate new unique reagents or materials.

Data and Code Availability:

- ERA5 and ERA5-Land data are publicly available from the Copernicus Climate Data Store (https://cds.climate.copernicus.eu). CERRA data are available from the same source. Climate Digital Twin data are available through Destination Earth (https://destination-earth.eu). ARA data are currently available on request and are scheduled to be publicly available by 2026. Analysis results generated in this study are available from the lead contact upon reasonable request.
- All code used for analysis and visualization is available at https://github.com/ischicker/perspectives_iScience_ClimateEnergyNexus_2025 and will be archived on Zenodo after manuscript acceptance.
- Any additional information required to reanalyse the data reported in this paper is available from the lead contact upon request.

## ACKNOWLEDGMENTS

This work was supported by the Digital Twin Austria project HectoRenew (FO999918390) and the Austrian Climate Research Programme (ACRP) project EnergyProtect(FO999901419). We thank colleagues and collaborators from the renewable energy sector for fruitful discussions on adverse and high impact weather and their needs and feedback on the manuscript. We acknowledge the use of ERA5, ERA5-Land, CERRA data from ECMWF, the Austrian ensemble reanalysis data ARA, and Destination Earth digital twin Climate DT data.